\documentclass[11pt,a4paper]{article}
\pdfoutput=1
\usepackage{graphicx}
\usepackage{jheppub}
\usepackage{amsfonts,amssymb}
\usepackage{amsmath}
\usepackage{epsfig}
\usepackage{bm}

\def\Re{{\rm Re}\,}
\def\Im{{\rm Im}\,}
\def\dLogT{\sigma}
\def\DLogT{\Sigma}

\begin{document}

\preprint{MIT-CTP/4602 \\ \phantom{.} \hfill PUPT-2473}

\title{Complexified boost invariance and holographic \\heavy ion collisions}

\author[a]{Steven S. Gubser}
\affiliation[a]{Joseph Henry Laboratories, Princeton University, Princeton, NJ 08544, USA}
\author[b]{and Wilke van der Schee}
\affiliation[b]{Center of Theoretical Physics, Massachusetts Institute of Technology, Cambridge, MA 02139, USA}

\date{\today}

\abstract{
At strong coupling holographic studies have shown that heavy ion collisions do not obey normal boost invariance. Here we study a modified boost invariance through a complex shift in time, and show that this leads to surprisingly good agreement with numerical holographic computations. When including perturbations the agreement becomes even better, both in the hydrodynamic and the far-from-equilibrium regime. One of the main advantages is an analytic formulation of the stress-energy tensor of the longitudinal dynamics of holographic heavy ion collisions.
}

\maketitle

\section{Introduction}
\noindent
In the study of heavy ion collisions it is often assumed that the collision is approximately boost invariant in the collision direction. This provides one of the simplest models of an expanding plasma. In weakly coupled QCD, collisions are indeed expected to be boost invariant over an expanding range of rapidities at asymptotically high energies. Nevertheless, at current experimental energies the rapidity distribution looks more like a Gaussian \cite{Afanasiev:2002mx, Bearden:2004yx, Back:2004je}, rather than the flat boost invariant distribution, which suggests current collisions are not yet in the weakly coupled regime.

One theoretical indication that heavy ion collisions are not (entirely) boost invariant comes from modelling these collisions in strongly coupled gauge theories, through their gravitational dual \cite{Chesler:2010bi, Casalderrey-Solana:2013aba,Chesler:2013lia,Casalderrey-Solana:2013sxa}. In this case the state before the collision is manifestly not boost invariant, and in fact the resulting thermalized plasma is not boost invariant. Instead, the thermalized plasma has a very constant temperature at constant real time (in the lab frame) \cite{Casalderrey-Solana:2013aba, vanderSchee:2014qwa}, as opposed to constant proper time expected from boost invariance.

In this paper we will try to link these two approaches by slightly modifying the definition of boost invariance, as done in \cite{Gubser:2012gy}. For this we look at boost invariant hydrodynamics and then shift the time variable by a complex parameter. We then look at the real part of the resulting stress tensor, which is manifestly conserved and in fact also described by hydrodynamics at late times.

The paper briefly reviews both the holographic model of heavy ion collisions  and the complexified boost invariance in section \ref{shocks} and \ref{complex}, after which we compare the resulting stress tensors and conclude that the agreement is surprisingly good. This has as a definite advantage that all computations using complexified boost invariance are completely analytic, while colliding heavy ions using holography requires advanced numerical techniques.

\section{Colliding planar shock waves in AdS\label{shocks}}
\noindent
To represent a heavy ion collision using the AdS/CFT duality we will use gravitational shock waves moving at the speed of light in AdS$_5$  \cite{Janik:2005zt,Albacete:2008vs,Gubser:2008pc,Grumiller:2008va,Aref'eva:2009wz}; in the CFT these correspond to lumps of energy moving at the speed of light, in this case in the strongly coupled, large-$N_c$ limit of $\mathcal{N}=4$ SU($N_c$) SYM, with $N_c$ the number of colors. These collisions hence do not directly model collisions in real-world QCD, but they nevertheless can give general insights of colliding lumps of energies in strongly coupled gauge theories (see \cite{Gubser:2009fc, CasalderreySolana:2011us, DeWolfe:2013cua} for reviews of AdS/CFT and heavy ion collisions).

In this study we furthermore restrict to planar symmetry in the transverse plane, such that solving the gravitational dynamics numerically is tractable \cite{Chesler:2010bi, Casalderrey-Solana:2013aba,Chesler:2013lia,Casalderrey-Solana:2013sxa, vanderSchee:2014qwa}. The initial condition for such a collision is then given by two single shocks, having all non-trivial dynamics in the `beam direction' ($z$), with an energy-momentum tensor whose only non-zero components are
\begin{equation}
T_{\pm \pm }(z_\pm) = \frac{N_{c}^{2}}{2 \pi^{2}} \, \frac{\mu^3}{\sqrt{2 \pi} w} \, e^{-z_\pm^2/2 w^2},
\end{equation}
where $z_\pm = t\pm z$, $w$ is the width of the sheets, $\mu^3$ is essentially the energy per transverse area and the sign depends on the direction of motion of each shock.

\begin{figure*}[t!]
  \centering
  \includegraphics[width=0.475\linewidth]{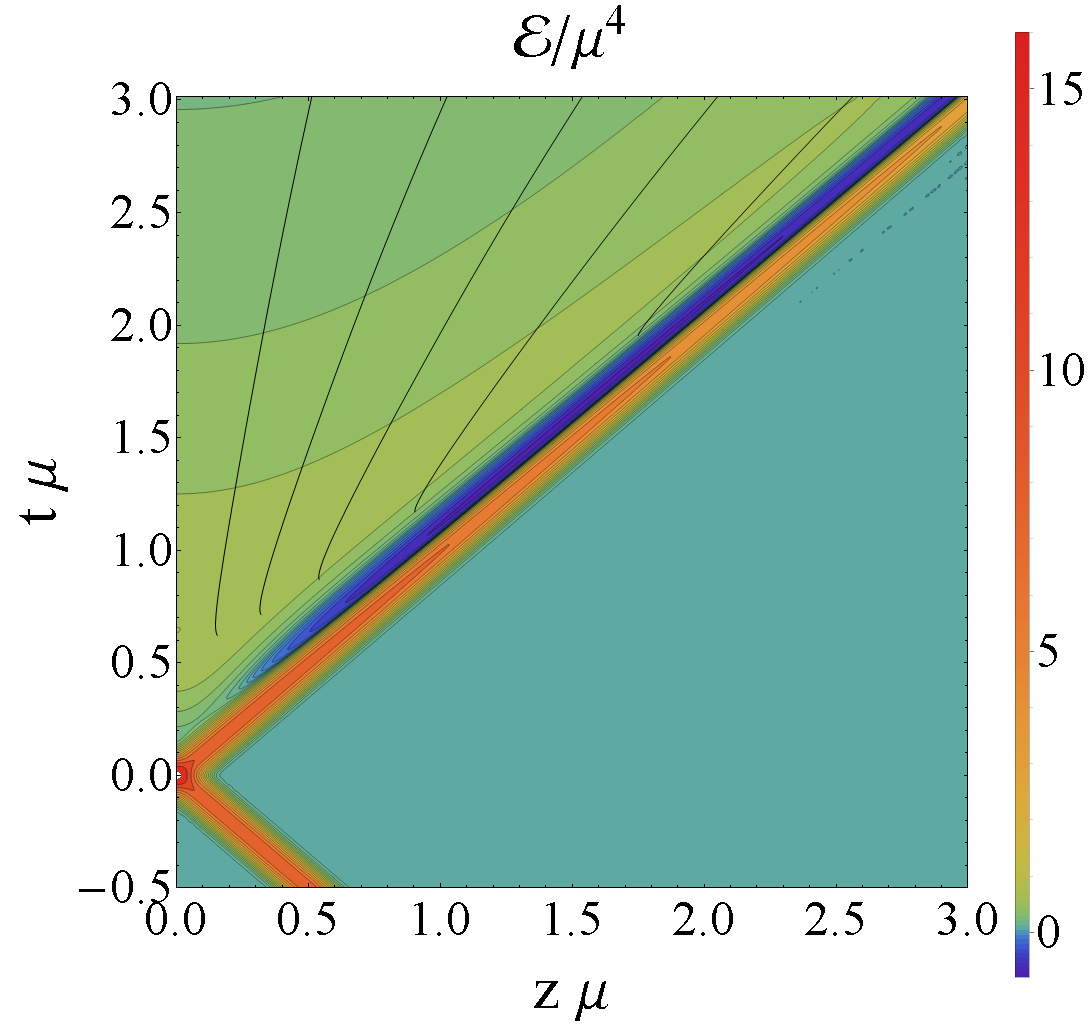}
  \includegraphics[width=0.45\linewidth]{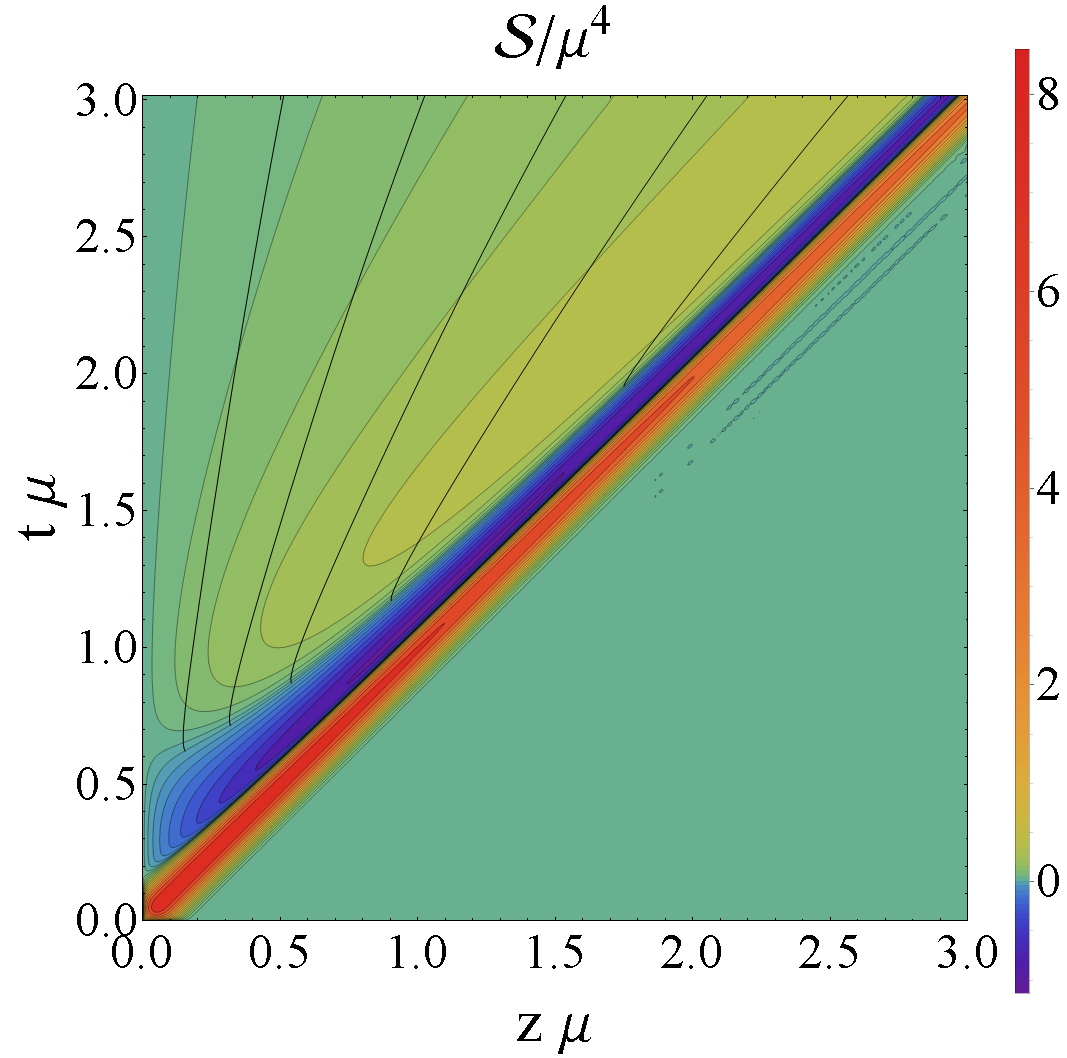}
  \caption{\label{fig:shocks} We plot the energy density  ($\mathcal{E} = \frac{2\pi^2}{N_c^2} T^t_t$) and flux ($\mathcal{S} = \frac{2\pi^2}{N_c^2} T^t_z$) after a collision of shock waves as computed in \cite{Casalderrey-Solana:2013aba}, with $\mu w = 0.05$. The black lines correspond to stream lines, stopping where a local velocity is not defined anymore \cite{Arnold:2014jva}. In the rest of the paper we will focus on snapshots at $\mu t = 1.5$ and $\mu t = 3$.}
\end{figure*}

This type of collision leads to rich physics in the field theory, especially when the dimensionless width $\mu w$ is small \cite{Casalderrey-Solana:2013aba}. First the shocks pass through each other virtually unperturbed, after which the original shocks slowly decay, leaving a plasma described by hydrodynamics in the middle (fig. \ref{fig:shocks}). Interestingly, there is a trailing far-from-equilibrium region behind the shocks where the energy density is negative, and no local rest frame can be defined, as recently expanded on in \cite{Arnold:2014jva}.

\subsection{Comparing with boost invariant hydrodynamics}

\begin{figure*}[t!]
  \centering
  \includegraphics[width=0.6\linewidth]{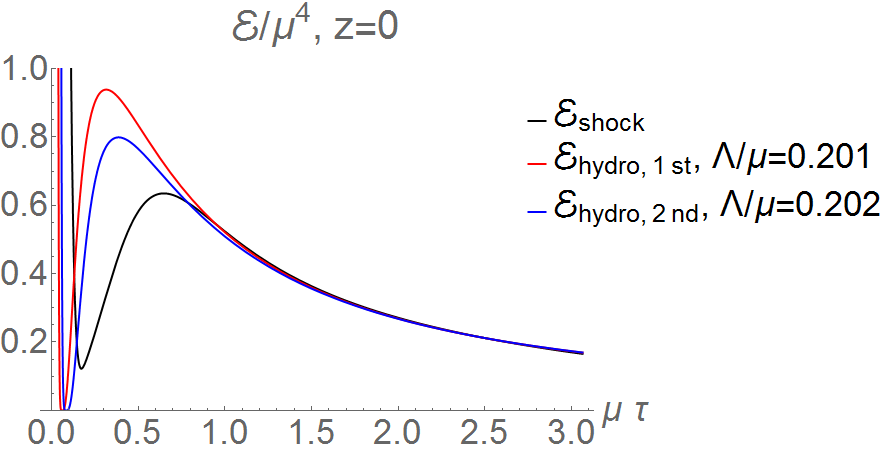}
  \caption{\label{fig:BIhydro} We plot the energy density  of the collision from figure \ref{fig:shocks} at mid-rapidity, i.e. $z=0$ (black). Also included are 1st and 2nd order hydrodynamic fits, from which we can extract that $\Lambda/\mu$ in eqn. \ref{eq:janik-heller} is 0.2. As already noted in \cite{Chesler:2013lia} it is interesting that the boost invariant fit at this fixed rapidity works very well for this limited time frame (later deviations will develop since the shock collisions are not boost invariant).}
\end{figure*}

The dynamics above can be compared with what one should find using boost invariant hydrodynamics. There, one expects all dynamics just to depend on proper time $\tau = \sqrt{t^2-z^2}$, whereby the stress tensor is completely determined by the local energy density given as \cite{Heller:2007qt}:
\begin{equation}\label{eq:janik-heller}
 \mathcal{E}_{BI}(\tau)=\frac{N_c^2 \Lambda^4}{2\pi^2}\left[
\frac{1}{(\Lambda \tau)^{4/3}}-\frac{2 \eta_0}{(\Lambda \tau)^{2}}+\frac{1}{(\Lambda \tau)^{8/3}} 
 \left(\frac{10}{3} \eta_0^2+ \frac{6 \ln
     2-17}{36\sqrt{3}}\right)+\mathcal{O}(\tau^{-10/3})\right],
\end{equation}
with $\eta_0=\frac{1}{\sqrt{2}\, 3^{3/4}}$ and $\Lambda$ the only parameter, setting the units in this equation. In this paper we will extract this parameter by demanding that the late time shock waves (around $\mu t = 3$ and at mid-rapidity, $z=0$) is well described by this second order boost invariant hydrodynamic formula. This gives us $\Lambda/\mu = 0.20$, as shown in figure \ref{fig:BIhydro}.

\begin{figure*}[t!]
  \centering
  \includegraphics[width=0.43\linewidth]{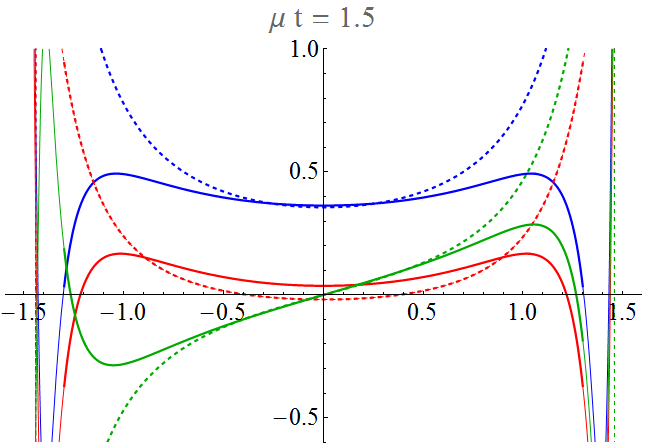}\,\,
  \includegraphics[width=0.52\linewidth]{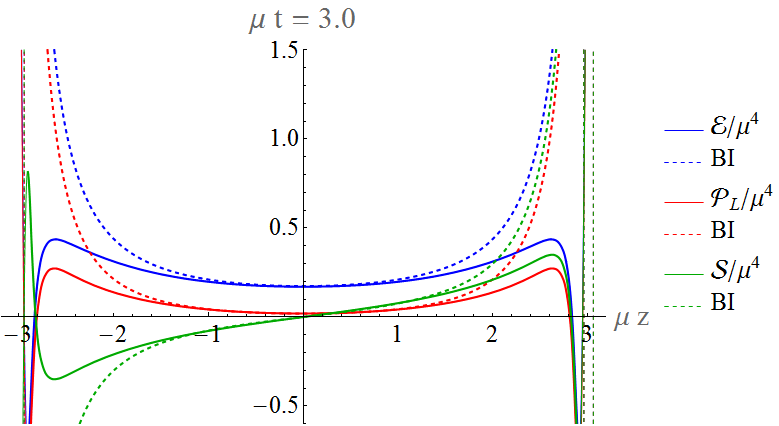}
  \caption{\label{fig:BIhydrosnap} In blue, red and green we respectively show the energy density , longitudinal pressure and flux of the shock collision at $\mu t=1.5$ (left) and $\mu t=3.0$ (right). The dashed lines show the boost invariant approximation (eqn.  \ref{eq:janik-heller}), which has $\Lambda$ being fitted at mid-rapidity, as displayed in fig. \ref{fig:BIhydro}. While boost invariance can give a reasonable description at mid-rapidity, there is a clear violation at higher rapidities. The curves are thick where we sample the difference between the approximated and full solution (i.e. having $\tau>\tau_{min}$) , in this case giving $\Delta =1.08$ (see eqn. \ref{eq:quantifier})}
\end{figure*}

\begin{figure*}[t!]
  \centering
  \includegraphics[width=0.55\linewidth]{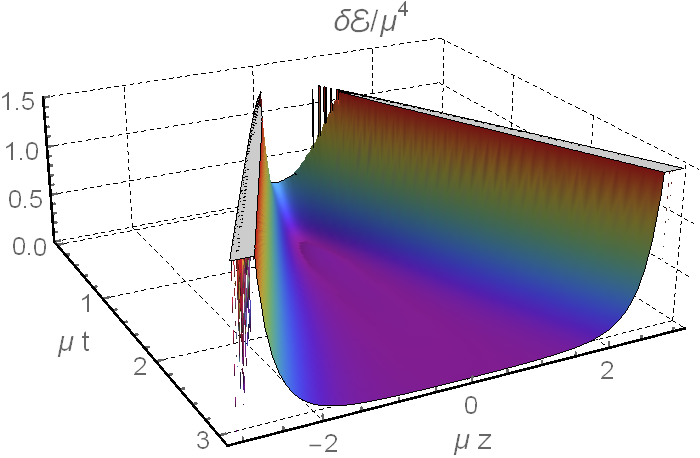}
  \caption{\label{fig:BIhydro3D} Here we plot the absolute difference in energy density of the boost invariant approximation with the full result as a function of time and longitudinal direction. It is clear that the approximation works nicely around mid-rapidity, while being violated badly at high rapidities. Figures for the pressures and flux look almost identical.}
\end{figure*}

We are now able to compare the full shock evolution with the boost invariant approximation. For this we chose to compare two snapshots in time ($\mu t=1.5$ and $\mu t=3.0$), where we restrict the comparison to the region where $\tau> \tau_{min}=0.75/\mu$, which includes part of the far-from-equilibrium evolution with i.e. negative energy densities, but does not include the lightcone with the original shocks. As a measure of how well we approximate the shocks we hence use the following dimensionless quantifier $\Delta$:
\begin{eqnarray}
\delta(t) &=& \int_{-\sqrt{t^2-\tau_{min}^2}}^{\sqrt{t^2-\tau_{min}^2}} dz \sqrt{\delta\mathcal{E}^2+\delta\mathcal{P}_L^2+\delta\mathcal{S}^2}\\
\Delta/\mu^3 &=& \delta(1.5/\mu) + \delta(3.0/\mu),\label{eq:quantifier}
\end{eqnarray}
where $\delta\mathcal{E} = \mathcal{E}_{shock} - \mathcal{E}_{BI}$, and $\delta\mathcal{P}_L$ and $\delta\mathcal{S}$ accordingly for the longitudinal pressure and flux. For boost invariant hydrodynamics (eqn. \ref{eq:janik-heller}) this then leads to $\Delta=1.08$, as illustrated in figures \ref{fig:BIhydrosnap} and \ref{fig:BIhydro3D}. 

\section{Complexified boost invariance\label{complex}}
\noindent
Complexified boost invariance refers to a scheme \cite{Gubser:2012gy,DeWolfe:2013cua} for producing solutions to the conservation equations $\nabla^\mu T_{\mu\nu} = 0$ by first formulating a complexified version of relativistic conformal hydrodynamics, in which the complexified stress tensor satisfies constitutive relations
 \begin{equation}\label{eq:TCrelations}
  T^{\bf C}_{\mu\nu} = \epsilon^{\bf C} u^{\bf C}_\mu u^{\bf C}_\nu + 
    {\epsilon^{\bf C} \over 3} (g_{\mu\nu} + u^{\bf C}_\mu u^{\bf C}_\nu)
 \end{equation}
with $\epsilon^{\bf C}$ and $u^{\bf C}_\mu$ also complex.  We impose the constraint $u^{\bf C}_\mu u_{\bf C}^\mu = -1$, and we insist that $T^{\bf C}_{\mu\nu}$ is conserved: $\nabla^\mu T^{\bf C}_{\mu\nu} = 0$.  Because the conservation equation is linear, the real part of $T^{\bf C}_{\mu\nu}$ is also conserved, and so we can set
 \begin{equation}\label{eq:RealPart}
  T_{\mu\nu} = \Re\{ T^{\bf C}_{\mu\nu} \} \,.
 \end{equation}
Intuitively, it makes sense that the complexified stress tensor should encode aspects of strongly dissipative dynamics; however, we have so far been unable to make this intuition more precise.  Our purpose here is to use the ansatz (\ref{eq:TCrelations})-(\ref{eq:RealPart}) to construct a variant of Bjorken flow which will compare favorably with the numerical treatment of thin shocks described in the previous section.

Bjorken flow can be constructed using the boost symmetry $B_3 = t \partial_z + z \partial_t$.  The proper time $\tau = \sqrt{t^2-z^2}$ is essentially the unique combination of $t$ and $z$ invariant under $B_3$.  (Essentially unique refers to the fact that functions only of $\tau$ are also invariant.)  One can summarize the Bjorken ansatz as
 \begin{equation}\label{eq:BjAnsatz}
  {\cal E} = {\cal E}(\tau) \qquad\qquad
  u_\mu = -\partial_\mu \tau \,,
 \end{equation}
where the normalization of $u_\mu$ is chosen so that $u_\mu u^\mu = -1$ with $u_t < 0$.  Note that $u_\mu$ is completely fixed at this point; ${\cal E}(\tau)$, on the other hand, must be fixed through the conservation equations $\nabla_\mu T^{\mu\nu} = 0$.

If we alter the boost generator $B_3$ to $B_3 + {\bf t}_3 T_3$ where $T_3 = \partial_z$ is the generator of translations in the $z$ direction, then we can see by direct calculation that the invariant combination of $t$ and $z$ is now $\sqrt{(t+{\bf t}_3)^2 - z^2}$.\footnote{It is slightly counterintuitive that shifting $B_3$ by a translation generator in the $z$ direction amounts to a shift $t \to t + {\bf t}_3$ of the time variable.  An easier example to visualize is a rotation around a point $y_0$ on the $y$ axis, which indeed acts as a combination of a rotation around the origin and a translation in the $x$ direction.}  For real ${\bf t}_3$, such a shift will lead to a trivial alteration of Bjorken flow, where the moment of impact becomes non-zero.  If ${\bf t}_3$ has an imaginary part, then we can proceed as outlined above to obtain first a complex conserved stress tensor, and then a real one.  That is, we set
 \begin{equation}\label{eq:BjC}
  {\cal E} = {\cal E}^{\bf C}(\tau^{\bf C}) \qquad\qquad
  u^{\bf C}_\mu = -\partial_\mu \tau^{\bf C} \qquad\qquad\hbox{where}\qquad
  \tau^{\bf C} = \sqrt{(t+{\bf t}_3)^2 - z^2} \,,
 \end{equation}
where as before $u^{\bf C}_\mu u_{\bf C}^\mu = -1$, and this condition uniquely fixes $u^{\bf C}_\mu$ up to an overall sign which may be fixed by demanding that $\Re\{u^{\bf C}_\mu\} < 0$ at $z=0$ for $t>0$: in short,
 \begin{equation}\label{eq:um}
  u^{\bf C}_\mu = 
    \left( -{t+{\bf t}_3 \over \sqrt{(t+{\bf t}_3)^2 - z^2}}, 0, 0, 
    {z \over \sqrt{(t+{\bf t}_3)^2 - z^2}} \right) \,.
 \end{equation}
For the conformal, inviscid stress tensor (\ref{eq:TCrelations}), the original Bjorken ansatz (\ref{eq:BjAnsatz}) leads to ${\cal E} = \hat{\cal E}_0/\tau^{4/3}$ for some constant $\hat{\cal E}_0$, and since all we have done is to translate the flow in time, it must be that
 \begin{equation}\label{eq:BjCE}
  {\cal E}^{\bf C} = {\hat{\cal E}^{\bf C}_0 \over (\tau^{\bf C})^{4/3}} \,,
 \end{equation}
where $\hat{\cal E}^{\bf C}_0$ is an integration constant.  For the same reason, a stress tensor that includes viscosity and higher order derivative terms will lead to a functional form for ${\cal E}^{\bf C}$ which is identical to the one obtained before translation: for example, the solution (\ref{eq:janik-heller}) carries over immediately to a complexified flow.  Note that this simple construction goes through when the equation of state is known analytically.  For example, it goes through for pressure $P = c_s^2 {\cal E}$ for any speed of sound $c_s^2$, because then $P$ can be found unambiguously even when ${\cal E}$ is complex.  If $P$ is known only for real values of ${\cal E}$, then in general it will not have any analytic continuation into the complex plane, and if it does, that continuation may have singularities which render the construction ill-defined.  Likewise any transport coefficients that enter must be capable of being evaluated for complex ${\cal E}$.

For large $\tau$ and fixed spacetime rapidity (defined as $\eta = \tanh^{-1} {z \over t}$), we see that $\tau^{\bf C} = \tau + {\bf t}_3 \cosh\eta + {\cal O}(1/\tau)$.  This implies that we recover ordinary Bjorken flow at late proper times, up to power law corrections.  For this to work, ${\cal E}^{\bf C}$ must have a positive real part.  It was observed in \cite{Gubser:2012gy} that when ${\bf t}_3$ is on the positive imaginary axis, there is only one choice of phase of $\hat{\cal E}^{\bf C}_0$ which leads to an inviscid flow (\ref{eq:TCrelations}) whose real part has a well-defined rest frame everywhere in the forward light-cone of the collision plane at $t=z=0$.  That choice is $\arg \hat{\cal E}^{\bf C}_0 = \pi/3$.  As discussed in \cite{Arnold:2014jva}, this positivity condition is not necessary; in light of the results of \cite{Casalderrey-Solana:2013aba}, it is not even desirable.  For $-\pi/2 < \arg \hat{\cal E}^{\bf C}_0 < \pi/3$, and at late times, one finds positive $T^{00}$ at mid-rapidity but negative $T^{00}$ sufficiently near the light-cone.  This statement holds true even in the presence of viscous and second-order corrections, which makes sense because such corrections generally become less important at late times.

\subsection{Second order hydrodynamics\label{sub:secondhydro}}

One can add terms to the complexified stress tensor (\ref{eq:TCrelations}) incorporating viscosity and higher order transport coefficients.  In the interests of a self-contained treatment, we will indicate how this is done through second order in derivatives, restricting to a flat background metric and zero vorticity.  A more comprehensive presentation can be found, for example, in \cite{Loganayagam:2008is}.  First we define the projection tensor
 \begin{equation}
  P^{\mu\nu} = g^{\mu\nu} + u^\mu u^\nu
 \end{equation}
so that the inviscid hydrodynamic stress tensor is simply
 \begin{equation}\label{idealhydro}
  T_{(0)}^{\mu\nu} = {\cal E} u^\mu u^\nu + {{\cal E} \over 3} P^{\mu\nu} \,,
 \end{equation}
where we set
 \begin{equation}
  {\cal E} = {3 N_c^2 \over 8\pi^2} (\pi T)^4
 \end{equation}
as appropriate for strongly coupled ${\cal N}=4$ super-Yang-Mills theory with $N_c$ colors.  Here and below, we omit all instances of ${\bf C}$, understanding that the established tensor structures in the literature are to be modified simply through replacing $u_\mu \to u_\mu^{\bf C}$ and $T \to T^{\bf C}$ everywhere.  The first derivative modification of the stress tensor is
 \begin{equation}\label{viscoushydro}
  T_{(1)}^{\mu\nu} = -2 \lambda \eta \sigma^{\mu\nu} \,,
 \end{equation}
where $\lambda$ is a formal parameter counting the number of derivatives (set to one in the end), and we have defined the shear tensor as
 \begin{equation}
  \sigma^{\mu\nu} = {1 \over 2} (P^{\mu\lambda} \partial_\lambda u^\nu + 
     P^{\nu\lambda} \partial_\lambda u^\mu) - 
     {1 \over 3} P^{\mu\nu} \partial_\lambda u^\lambda \,,
 \end{equation}
and, for strongly coupled ${\cal N}=4$ super-Yang-Mills theory,
 \begin{equation}
  \eta = {N_c^2 \over 8\pi^2} (\pi T)^3 \,.
 \end{equation}
The second derivative modification of the stress tensor is simplified when we specialize to gradient flow, where the vorticity vanishes: then
 \begin{align}
  T^{(2)}_{\mu\nu} &= 2 \lambda^2 \eta \tau_\pi u^\lambda {\cal D}_\lambda \sigma^{\mu\nu} + 
    \lambda^2 \xi_\sigma \left[ \sigma^\mu{}_\lambda \sigma^{\lambda\nu} - 
      {P^{\mu\nu} \over 3} \sigma^{\alpha\beta} \sigma_{\alpha\beta} \right]
 \end{align}
where ${\cal D}_\lambda \sigma^{\mu\nu}$ is a covariant derivative of the shear tensor which transforms tensorially under conformal transformations, specified by the following definitions:
 \begin{align}
  {\cal D}_\lambda \sigma^{\mu\nu} &= \partial_\lambda \sigma^{\mu\nu} + 
    3 A_\lambda \sigma^{\mu\nu} + A_{\lambda\rho}^\mu \sigma^{\rho\nu} + 
      A_{\lambda\rho}^\nu \sigma^{\rho\mu} \\
  A_{\lambda\rho}^\mu &= g_{\lambda\rho} A^\mu - \delta_\lambda^\mu A_\rho - 
    \delta_\rho^\mu A_\lambda \\
  A_\mu &= u^\nu \partial_\nu u_\mu - {\partial_\nu u^\nu \over 3} u_\mu \,.
 \end{align}
Also, for strongly coupled ${\cal N}=4$ super-Yang-Mills theory, we have the parameters
 \begin{equation}
  \tau_\pi = {2-\log 2 \over 2\pi T} \qquad\qquad
  \xi_\sigma = {2\eta \over \pi T} \,.
 \end{equation}

\begin{figure*}[t!]
  \centering
  \includegraphics[width=0.43\linewidth]{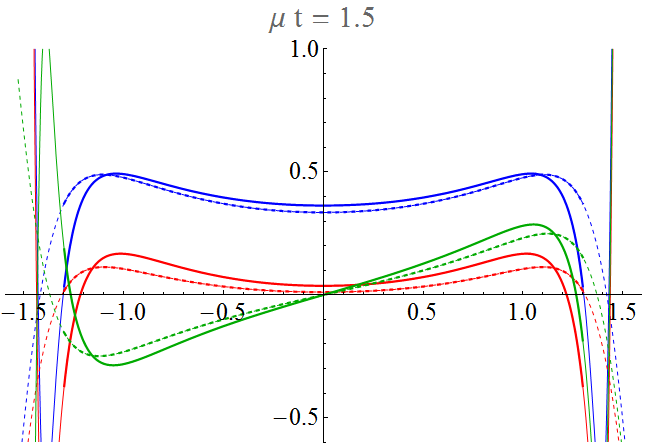}\,\,
  \includegraphics[width=0.52\linewidth]{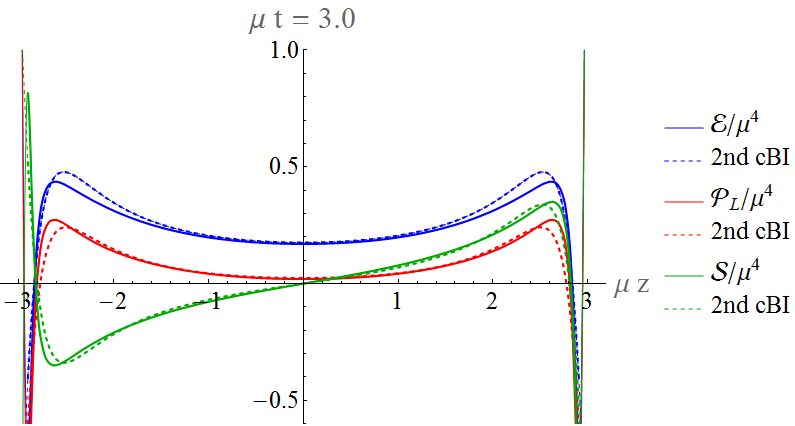}
  \caption{\label{fig:cBIhydrosnap} Here we plot the analog of figure \ref{fig:BIhydrosnap}, using complexified boost invariant second order hydrodynamics. For this we used $\mu {\bf t}_3 = 0.080+0.318 i$ and a phase of $\theta =-0.425$, leading to an accuracy of $\Delta = 0.106$ (see eqn. \ref{eq:quantifier}). First order hydrodynamics would only lead to slight differences ($\mu {\bf t}_3 = 0.084+0.326 i$, $\theta =-0.390$ and $\Delta = 0.102$).}
\end{figure*}

\begin{figure*}[t!]
  \centering
  \includegraphics[width=0.55\linewidth]{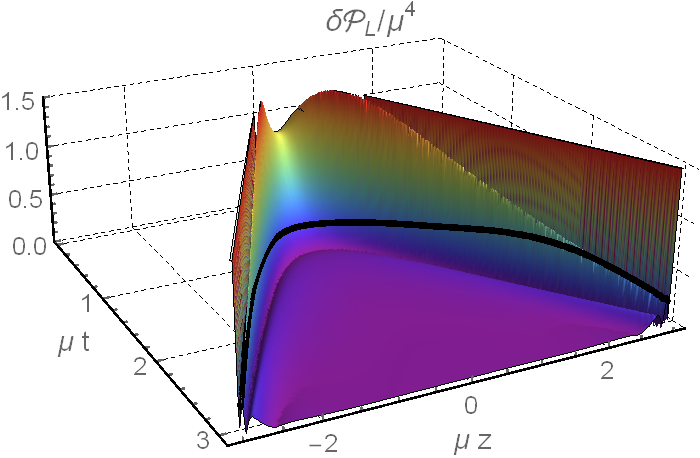}
  \caption{\label{fig:1cBIhydro3D}Here we plot the analog of figure \ref{fig:BIhydro3D} for second order viscous complexified boost invariant hydrodynamics for the parameters of figure \ref{fig:cBIhydrosnap}. The black line indicates the $\tau=\tau_{min}$ surface. Clearly using complexified boost invariance provides a good description over a much larger range of rapidities. We now plot the difference in longitudinal pressure, $\mathcal{P}_L =  \frac{2\pi^2}{N_c^2} T^z_z$, where again the energy and flux plots look very similar.}
\end{figure*}

After adding up the inviscid, first order and second order contributions we can then solve the conservation equations to indeed find $\cal E(\tau)$ as given in eqn. \ref{eq:janik-heller}.  We fixed $\Lambda$ by demanding that the leading late time asymptotics matches the fit of \ref{eq:janik-heller} presented in figure \ref{fig:BIhydro}.  Apart from $\Lambda$ this now has the complex ${\bf t}_3$ as two new degrees of freedom\footnote{The real part of ${\bf t}_3$ is of course a shift in real time, and it is therefore possible to argue that ${\bf t}_3$  should be purely imaginary. Here we chose to keep ${\bf t}_3$  arbitrary, but we found the real part to be negligibly small.}. Also, we can change the phase of either $\hat{\cal E}^{\bf C}_0$ or $T_{\mu\nu}$ before taking the real part, which we incorporated here by leaving $\hat{\cal E}^{\bf C}_0$ real and positive and multiplying $T_{\mu\nu}$ by  $e^{i\theta}$ before taking the real part. This leads to three parameters, which we find by minimizing $\Delta$ in eqn. \ref{eq:quantifier}. This way both first and second order hydrodynamics agree quantitatively well with the thin shocks of the previous section, having a $\Delta$ of ten times smaller, as shown in figures \ref{fig:cBIhydrosnap} and \ref{fig:1cBIhydro3D}.

\subsection{Including small perturbations}
\label{sec:Perturbed}

To discuss perturbations, a helpful preliminary is to consider much simpler dynamics than linearized hydrodynamics, namely the massless scalar in $1+1$ dimensions.  In this case, the general solution is of course $\phi = f(t+z) + g(t-z)$.  We may in particular consider the solution where
 \begin{equation}\label{WavePackets}
  \phi = (x_{\bf C}^+)^n \qquad\hbox{where}\qquad x_{\bf C}^\pm = t+{\bf t}_3 \pm z \,.
 \end{equation}
Here the parameter $n$ is the quantum number of $SO(1,1)_{\bf C}$ carried by the solution: This is meant in the sense that $(B_3 + {\bf t}_3 T_3) \phi = n \phi$.  Alternatively, $\phi = (x_{\bf C}^-)^{-n}$ carries the same quantum number.  In order to have a bounded solution, we should choose $(x_{\bf C}^+)^n$ when $n$ has negative real part and $(x_{\bf C}^-)^{-n}$ when $n$ has a positive real part.  These solutions describe wave-packets whose direction of motion is to the left if $\Re n < 0$ and to the right if $\Re n > 0$; more precisely, if $|\Re n|$ is large, they are wave-packets with momentum centered on $k = {\Re n \over \Im {\bf t}_3}$ whose width in momentum space is $\DLogT k \sim {\sqrt{|\Re n|} \over \Im {\bf t}_3}$.  A disadvantage of the solutions (\ref{WavePackets}) as compared to standard plane waves is that they do not form an orthonormal basis, so it is not so clear how one would expand an arbitrary solution as a superposition of solutions with definite quantum numbers under $SO(1,1)_{\bf C}$.  Linearized complexified hydrodynamics will have the same problem.  The level of our presentation, therefore, will be exploratory, where we consider perturbations with particular values of $n$ and inquire to what extent they improve the fit to numerical results.

Let us next treat the case of inviscid hydrodynamics.  It is helpful to define
 \begin{equation}
  g = x_{\bf C}^+ x_{\bf C}^- = (\tau^{\bf C})^2 \,,
 \end{equation}
and to analyze perturbations to linearized order starting from the ansatz
 \begin{equation}\label{PerturbAnsatz}
  u^\pm \equiv u^t \pm u^z = {x_{\bf C}^\pm \over \sqrt{g}} (1 \pm \nu) \qquad\qquad
  T = T_0(g) (1+\dLogT) \,,
 \end{equation}
where $T$ is the temperature and $T_0(g) = \hat{T}_0 / g^{1/6}$  is the unperturbed inviscid Bjorken flow solution  (with $\hat{T}_0=\sqrt{2} \Lambda ^{2/3}/\sqrt[4]{3} \pi$ from the ideal part of eqn. \ref{eq:janik-heller}).  We assume no perturbations in the transverse directions.  All quantities at this stage are understood to be complexified, with the real part to be taken at the end of the calculation, but we omit subscripted or superscripted ${\bf C}$s for brevity.  The perturbations $\nu$ and $\dLogT$ will be treated to linear order.  We are able to parametrize the general longitudinal perturbation as indicated in (\ref{PerturbAnsatz}) because $u^\pm$ are constrained to satisfy $u^+ u^- = 1$.  Assuming further
 \begin{equation}\label{SOcontent}
  \dLogT = \dLogT_n(t,z) \equiv (x_{\bf C}^+)^n \DLogT(g) \qquad\qquad
  \nu = \nu_n(t,z) \equiv (x_{\bf C}^+)^n {\rm N}(g) \,,
 \end{equation}
one finds that the ansatz (\ref{PerturbAnsatz}) satisfies the conservation equations for the full complexified stress tensor to linear order provided
 \begin{equation}\label{dvEoms}
  \DLogT' + {n \over 2} \DLogT + {n \over 6} {\rm N} = 0 \qquad\qquad
  {\rm N}' + {n \over 2} \DLogT + \left( {n \over 2} + {1 \over 3} \right) 
    {\rm N} = 0
 \end{equation}
where primes denote $d/d\log g$.  If instead one assumes $\dLogT = (x_{\bf C}^-)^n \DLogT(g) $ and $\nu = -(x_{\bf C}^-)^n {\rm N}(g)$ (note the explicit minus sign on $\nu$), then one obtains the same equations (\ref{dvEoms}).  One may straightforwardly solve (\ref{dvEoms}) to obtain
 \begin{align}
  \DLogT &= g^{-{1 \over 6} - {n \over 2}} \Bigg( C_1 \left[ 
    \left( 1 + \sqrt{1+3n^2} \right) g^{{1 \over 6} \sqrt{1+3n^2}} + 
    \left( -1 + \sqrt{1+3n^2} \right) g^{-{1 \over 6} \sqrt{1+3n^2}} \right]  \nonumber \\
   &\qquad\qquad\quad{} + 
    n C_2 \left[ g^{{1 \over 6} \sqrt{1+3n^2}} - g^{-{1 \over 6} \sqrt{1+3n^2}}
      \right] \Bigg)  \nonumber \\
  {\rm N} &= g^{-{1 \over 6} - {n \over 2}} \Bigg( -3n C_1 \left[
    g^{{1 \over 6} \sqrt{1+3n^2}} - g^{-{1 \over 6} \sqrt{1+3n^2}} \right]  \nonumber \\
   &\qquad\qquad\quad{} + C_2 \left[ 
    \left( -1 + \sqrt{1+3n^2} \right) g^{{1 \over 6} \sqrt{1+3n^2}} + 
    \left( 1 + \sqrt{1+3n^2} \right) g^{-{1 \over 6} \sqrt{1+3n^2}} \right] \Bigg) \,.
    \label{SoundModes}
 \end{align}
To actually assemble a physically relevant perturbation, our procedure is to start with a perturbation with definite weight $n$ under $SO(1,1)_{\bf C}$---namely a perturbation of the form (\ref{SOcontent})---and symmetrize it as follows:
 \begin{equation}\label{PertSym}
  \dLogT = \dLogT_n(t,z) + \dLogT_n(t,-z) \qquad\qquad
  \nu = \nu_n(t,z) - \nu_n(t,-z) \,.
 \end{equation}
Using $\DLogT$ and ${\rm N}$ given as in (\ref{SoundModes}) leads to a solution of the linearized equations.  Note that the minus signs in the expression (\ref{PertSym}) for $\nu$ are forced on us by the explicit overall sign on $\nu$ in the discussion below (\ref{dvEoms}).  More physically, a temperature perturbation that is even under $z \to -z$ must be accompanied by an odd perturbation in the velocity field.

\begin{table*}[!ht]
\caption{\label{Table} For a more quantitative comparison this table summarises $\delta(t)$ at different times, for boost-invariant, viscous complexified boost invariance and perturbed complexified boost invariance, all for two different $\tau_{min}$. As can also be seen in the figures it is clear that the perturbed complexified boost invariance leads to the best fit, especially at later times, and also that a lower $\tau_{min}$ leads to a significantly larger $\delta(t)$, indicating that most of the difference is located close to the lightcone where the original shocks are located.}
\begin{center}
\begin{tabular}{|l||c|c|c|c|c||c|c|c|c|c|c|c|c|c|c|c|c|} 
\hline
&
 \multicolumn{5}{c||} {$\mu \tau_{min}=0.375$} &  \multicolumn{5}{c|} {$\mu \tau_{min}=0.75$}
\\
\hline
\text{$\mu t$} & 1.0 & 1.5 & 2.0 & 2.5 & 3.0  & 1.0 & 1.5 & 2.0 & 2.5 & 3.0 \\
 \hline
$\delta_{\text{BI}} (t)$ &   0.30 & 0.50 & 0.73 & 1.0 & 1.3 & 0.061 & 0.21 & 0.40 & 0.63 & 0.87 \\
\hline
$\delta_{\text{2nd}} (t)$ &  0.18 & 0.15 & 0.12 & 0.11 & 0.12 & 0.028 & 0.049 & 0.048 & 0.047 & 0.057 \\
\hline
$\delta_{\text{pert}} (t)$ & 0.20 & 0.16 & 0.12 & 0.095 & 0.088 & 0.023 & 0.048 & 0.045 & 0.034 & 0.032 \\
\hline
\end{tabular}
\end{center}
\label{table}
\vspace{-0.5cm}
\end{table*}

The equations (\ref{dvEoms}) and the solutions (\ref{SoundModes}) can be improved by the inclusion of viscosity.  Specifically, one finds
 \begin{align}\label{ViscousDiffEQ}
  \DLogT' + {n \over 2} \DLogT + {n \over 6} {\rm N} + 
        {\lambda \over 18\pi g^{1/3} \hat{T}_0} \left[ \DLogT - 2n {\rm N} \right] &= 0  \\
  {\rm N}' + {n \over 2} \DLogT + \left( {n \over 2} + {1 \over 3} \right) {\rm N} - 
        {\lambda \over 6\pi g^{1/3}  \hat{T}_0} \left[ 2n \DLogT + (n^2-2) {\rm N} \right] &= 0 \,,
 \end{align}
where now of course the function $T_0(g)$ also includes the viscous term of eqn. \ref{eq:janik-heller}. In writing (\ref{ViscousDiffEQ}) we have systematically neglected terms at $O(\lambda^2)$ and higher, which in boost invariant hydro amounts to neglecting terms of order $g^{-2/3}=\tau^{-4/3}$.  Correspondingly, when constructing viscous improvements of the solutions (\ref{SoundModes}), we keep terms only through $O(\lambda)$: in other words, we expand 
 \begin{equation}\label{DNExpand}
  \DLogT = \DLogT_0 + \lambda \DLogT_1 \qquad\qquad
  {\rm N} = {\rm N}_0 + \lambda {\rm N}_1
 \end{equation}
and employ the expressions (\ref{SoundModes}) for $\DLogT_0$ and ${\rm N}_0$; then we extract ${\rm N}_1$ by plugging (\ref{DNExpand}) into (\ref{ViscousDiffEQ}) and dropping terms quadratic in $\lambda$.  The resulting expressions for $\DLogT_1$ and ${\rm N}_1$ are sums of fractional powers of $g$, just as in (\ref{SoundModes}) but somewhat more complicated. Note in particular that we can ignore the integration constants for ${\rm \Sigma}_1$ and ${\rm N}_1$, as they correspond to shifts in $C_1$ and $C_2$.

\begin{figure*}[t!]
  \centering
  \includegraphics[width=0.42\linewidth]{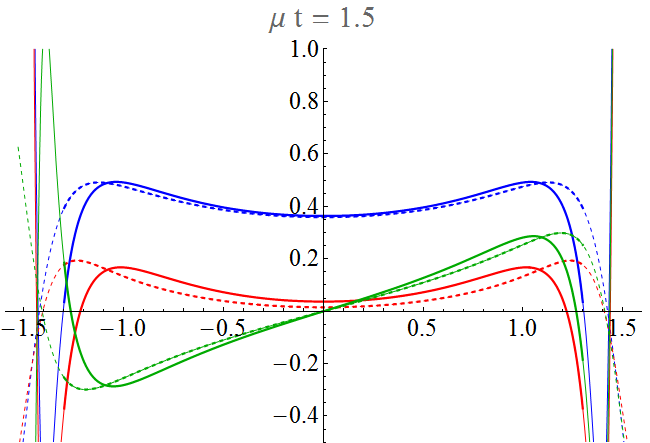}\,\,
  \includegraphics[width=0.56\linewidth]{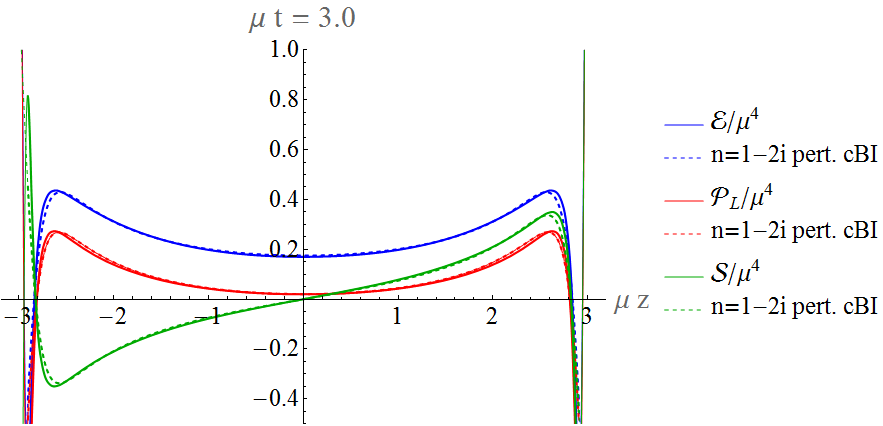}
  \caption{\label{fig:pertcBIhydrosnap} Here we plot the analog of figure \ref{fig:BIhydrosnap}, using perturbed first order viscous complexified boost invariant hydrodynamics, with $n=1-2i$ as representation of the perturbation. This leads to a significantly better fit, giving $\Delta = 0.079$, but at the price of having four extra (real) parameters. We interpret the fact that a perturbation with four extra parameters only improves the fit by about 25\% as an extra indication that the agreement found in figure \ref{fig:cBIhydrosnap} is indeed remarkable.}
\end{figure*}

\begin{figure*}[t!]
  \centering
  \includegraphics[width=0.55\linewidth]{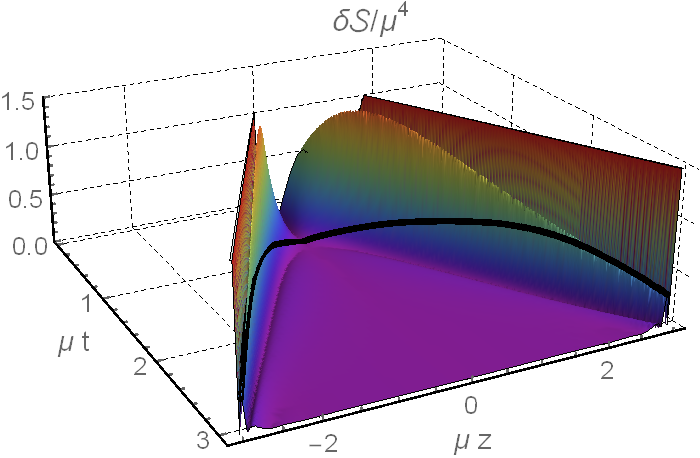}
  \caption{\label{fig:pertcBIhydro3D}Lastly we show the (absolute) difference in flux when using complexified boost invariance including small perturbations on top. The fit is about 25\%  better than without using perturbations, and again we note that the energy density and pressure behave very similarly.}
\end{figure*}

In order to now compare these perturbed solutions we started with the first order solution obtained in subsection \ref{sub:secondhydro}, including the ${\bf t}_3$ and $\theta$ found there. We then added a single perturbation as described above, giving us a modified velocity and temperature as in eqn. (\ref{PerturbAnsatz}), which in turn gives us through eqn.  (\ref{idealhydro}) and (\ref{viscoushydro}) the stress tensor which we ultimately compare with the numerics of the shock waves. For this comparison we tried $n=a + bi$ with $a$ and $|b|$ positive integers smaller than 5, again fitting the constants (in this case $C_1$ and $C_2$) to minimize $\Delta$ (eqn. (\ref{eq:quantifier})). This gave the best fit for $n=1-2i$ with an improvement of about 25\%, as is shown in figures \ref{fig:pertcBIhydrosnap} and \ref{fig:pertcBIhydro3D}. This comparison is illustrated more quantitatively in table \ref{Table}, where we also include a comparison at a different $\tau_{min}$.

Clearly with the extra degrees of freedom this improved fit is unsurprising and in that light the improvement is only modest. This could mean that we should add several perturbations (with even more degrees of freedom), or that the linearized approximation in $\lambda$ did not work well in the time range of interest. On the other hand, the fact that including five extra degrees of freedom improves the fit by only 25\% can be taken as an indication that the original fits found in subsection \ref{sub:secondhydro} are remarkably good.

\section{Discussion}
\noindent
This paper applied the idea of complexified boost invariance \cite{Gubser:2012gy} to strongly coupled heavy ion collisions, here holographically modelled as a collision of planar gravitational shock waves \cite{Casalderrey-Solana:2013aba}. By shifting time by a complex parameter we formally kept boost invariance as a symmetry in complexified coordinates, but the real part of the stress tensor has boost invariance explicitly broken. In this way we can keep on using analytic results relying on boost invariance, whereas the real part of the stress tensor can describe a wider range of physics. This allows in particular to compare with the plasma formed after a holographic heavy ion collision, which is manifestly not boost invariant.

The fits of subsection \ref{sub:secondhydro} by using 2nd order hydrodynamic complexified boost invariance are impressive; they involve only three parameters (the complex shift ${\bf t}_3$ and the phase angle $\theta$), and yet they fit the full stress energy tensor (with three independent functions) surprisingly well, over a large time domain, and a space domain extending all the way into far-from-equilibrium regions. When including perturbations on top of these profiles the quality of the fits improve by another 25\%, albeit at the price of extra parameters (the representation $n$, and the complex constants $C_1$ and $C_2$). Importantly, while the stress tensor from complex boost invariance is not necessarily described by (viscous) hydrodynamics, the agreement with the thin shocks shows that in the cases presented complex boost invariance does in fact allow a hydrodynamic description provided one is not too close to the lightcone.

Especially in light of the seemingly different results for longitudinal dynamics in weakly coupled \cite{Gelis:2010nm, Schenke:2010nt} and strongly coupled \cite{Casalderrey-Solana:2013sxa, vanderSchee:2014qwa} theories the further study of the rapidity distribution in heavy ion collisions will be of crucial importance. We therefore stress once more that all complexified boost invariant results here are completely analytic, whereas the gravitational shock wave collisions require relatively intensive numerics. Therefore, at a practical level, this analytic treatment may provide a useful simplified model of the implications of strong coupling for the longitudinal dynamics of the initial stage of a heavy ion collision. Last we note that the perturbation analysis of section~\ref{sec:Perturbed} is also completely analytic and could have useful applications to describe for instance thermal fluctuations in real heavy ion collisions \cite{Springer:2012iz}.

\section*{Acknowledgments}

We thank Jorge Casalderrey and Michal Heller for discussions. The work of S.S.G.~was supported in part by the Department of Energy under Grant No.~DE-FG02-91ER40671. WS is supported by the U.S. Department of Energy under grant Contract Number DE-SC0011090 and by a Utrecht University Foundations of Science grant.

\bibliographystyle{apsrev} \bibliography{draft}

\end{document}